\documentclass{iaus}
\usepackage{graphicx}

\def\fe{{\it Fermi}}
\def\gsim{ \lower .75ex \hbox{$\sim$} \llap{\raise .27ex \hbox{$>$}} } 
\def\lsim{ \lower .75ex\hbox{$\sim$} \llap{\raise .27ex \hbox{$<$}} }

\def\ama{$E_{\rm peak}-E_{\rm iso}$}
\def\yone{$E_{\rm peak}-L_{\rm iso}$}
\def\yonet{$E_{\rm peak}^{t}-L_{\rm iso}^{t}$}

\def\fermi{{\it Fermi}}

\def\ep{$E_{\rm peak}$}
\def\epo{$E^{\rm obs}_{\rm peak}$}

\def\eiso{$E_{\rm iso}$}
\def\liso{$L_{\rm iso}$}
\def\eg{$E_{\rm \gamma}$}

\title[Gamma Ray Bursts] 
{Gamma Ray Bursts Spectral--Energy correlations: recent results}

\author[Giancarlo Ghirlanda]   
{Giancarlo Ghirlanda
}
\affiliation{Osservatorio Astronomico di Brera, Via Bianchi 46 Merate I--23807 Italy
\\ email: {\tt giancarlo.ghirlanda@brera.inaf.it} }

\pubyear{2010}
\volume{275}  
\pagerange{1--8}
\jname{Jets at all Scales}
\editors{G.E. Romero, R.A. Sunyaev \& T. Belloni}
\begin{document}

\maketitle

\begin{abstract}

The correlations between the rest frame peak of the $\nu F_{\nu}$ spectrum of GRBs (\ep) and their 
isotropic energy (\eiso) or luminosity (\liso) could have several implications for the understanding of 
the GRB prompt emission. These correlations are presently founded on the {\it time--averaged} 
spectral properties of a sample of 95 bursts, with measured redshifts, collected by different instruments 
in the last 13 years (pre--\fe). One still open issue is wether these correlations have a physical origin or are due to instrumental 
selection effects. By studying 10 long  and 14 short GRBs detected by \fe\ we find that a strong {\it time--resolved} correlation 
between \ep\ and the luminosity \liso\ is present within individual GRBs and that it
is consistent with the {\it time--integrated} correlation. This result is a direct proof of the existence in both short and 
long GRBs of a similar physical link between the hardness and the luminosity which is not due to instrumental selection effects. 
The origin of the \yone\ correlation should be searched in the radiation mechanism of the prompt emission. 
\end{abstract}


\section{Spectral--Energy correlations}
One of the key properties of the prompt emission of gamma ray bursts
(GRBs) that is still poorly understood concerns the spectral--energy
correlations found when considering the time--integrated spectra of bursts
of known redshift. The peak energy of the spectrum, \ep\ in the $\nu F_{\nu}$ representation, is 
strongly correlated with the isotropic luminosity \liso\ (Yonetoku et al. 2004)  or with the isotropic 
energy \eiso\ (Amati et al. 2002), and more tightly with the collimation--corrected energy \eg\
(Ghirlanda, Ghisellini \& Lazzati 2004). 

There are two strong motivations for studying these correlations: understand their  
physics and use them to standardize the GRB energetics, making them cosmological tools.

\begin{figure}
\vspace*{-0.5 cm}
\hspace*{-0.8 cm}
\includegraphics[width=2.5in, height=5.8in,angle=90]{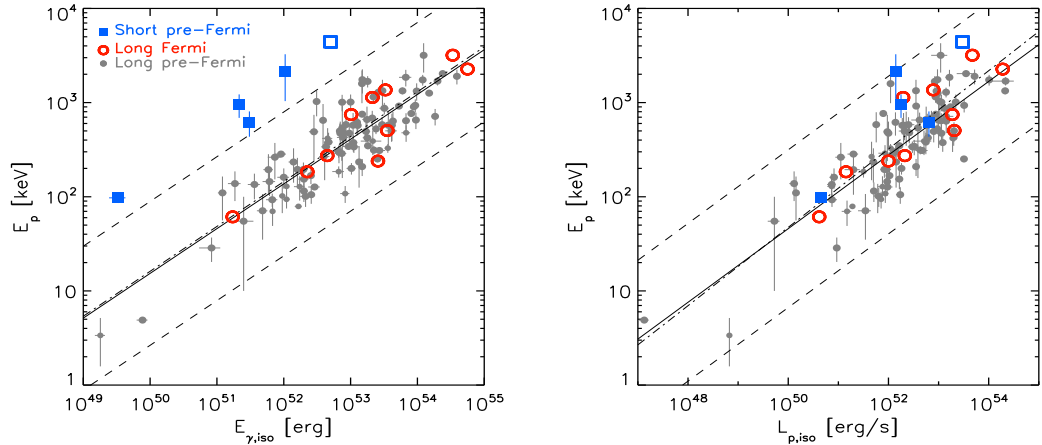} 
\vspace*{-0.6 cm}
\caption{Rest frame peak energy versus isotropic energy (left) and luminosity (right) of 95 long GRBs detected before 
\fe, i.e. pre--\fe\ sample (filled grey circles). 
The 10 \fe\ long GRBs with measured redshift and \epo\ (up to July 2009) are shown by the open (red) circles, respectively. 
The only short \fe\ GRB with measured redshift and measured \epo\ (090510)
is shown by the open (blue) square. The filled (blue) squares are the 4 short GRBs with measured redshift and \epo\ 
detected by other instruments than \fe. The solid line is the best fit to the pre-\fe\ sample, 
while the dashed lines represent its 3$\sigma$ scatter. The dot--dashed line 
is the best fit including the 10 \fe\ long GRBs. (Figure adapted from Ghirlanda et al. 2009, 2010, 2010a)}
\label{fig1}
\end{figure}

However, the \ama\ and \yone\ correlations have been derived considering long GRBs (Fig.~\ref{fig1} filled grey and 
open red circles) due to the lack of measured redshifts of short GRBs. It has been shown recently 
(Ghirlanda et al. 2009) that the few short GRBs with measured $z$ (filled blue squares in Fig.~\ref{fig1}) do not follow 
the \ama\ correlation (Fig.~\ref{fig1}, left panel) but they are consistent with the \yone\ correlation (Fig.~\ref{fig1}, right panel). 

A still debated issue is wether these correlations have a physical origin or 
they are the result of instrumental selection effects (Nakar \& Piran 2005; Band \& Preece 2005; Butler et al. 2007, 
Butler, Kocevski \& Bloom 2009; Shahmoradi \& Nemiroff 2009 but see Ghirlanda et al. 2005, Bosnjak et al. 2008, 
Ghirlanda et al. 2008; Nava et al., 2008; Krimm et al. 2009; Amati, Frontera \& Guidorzi  2009).

A completely orthogonal possibility (with respect to the debate opened in the literature - e.g. Butler, Kocevski \& Bloom 2009 
but see e.g. Ghirlanda et al. 2008) to answer this question is to study {\it individual} bursts to see whether the luminosity 
and peak energy at different times during the prompt phase correlate. If they do, and furthermore 
if the slope of this {\it time--resolved} correlation (indicated \yonet\ hereafter) is similar to the {\it time--integrated} 
\yone\ correlation found among different bursts, then we should conclude that the spectral energy correlations
are surely a manifestation of the physics of GRBs and not the result of instrumental selection effects.

The two questions we want to answer are: (1) is there a time resolved spectral energy correlation within individual GRBs? and
(2) is this correlation present in both short and long GRBs? 

To answer these questions the {\it time--resolved} spectral analysis of GRB is required. Moreover, in order to follow the 
evolution of the spectrum and its peak energy \ep\ in time all over the burst duration, a large spectral energy window 
is desirable like that of the Gamma Burst Monitor (GBM, 8keV-40MeV) onboard the \fe\ satellite. 
Here we present the main results of the study of the spectral evolution of 10 long GRBs and 14 short GRBs detected 
by \fe\ (these results have been published in Ghirlanda et al. 2010, 2010a). 

\section{Time resolved spectral--energy correlations}

\begin{figure}
\vspace*{-0.5 cm}
\hspace*{-0.8 cm}
\includegraphics[width=2.8in, height=2.7in]{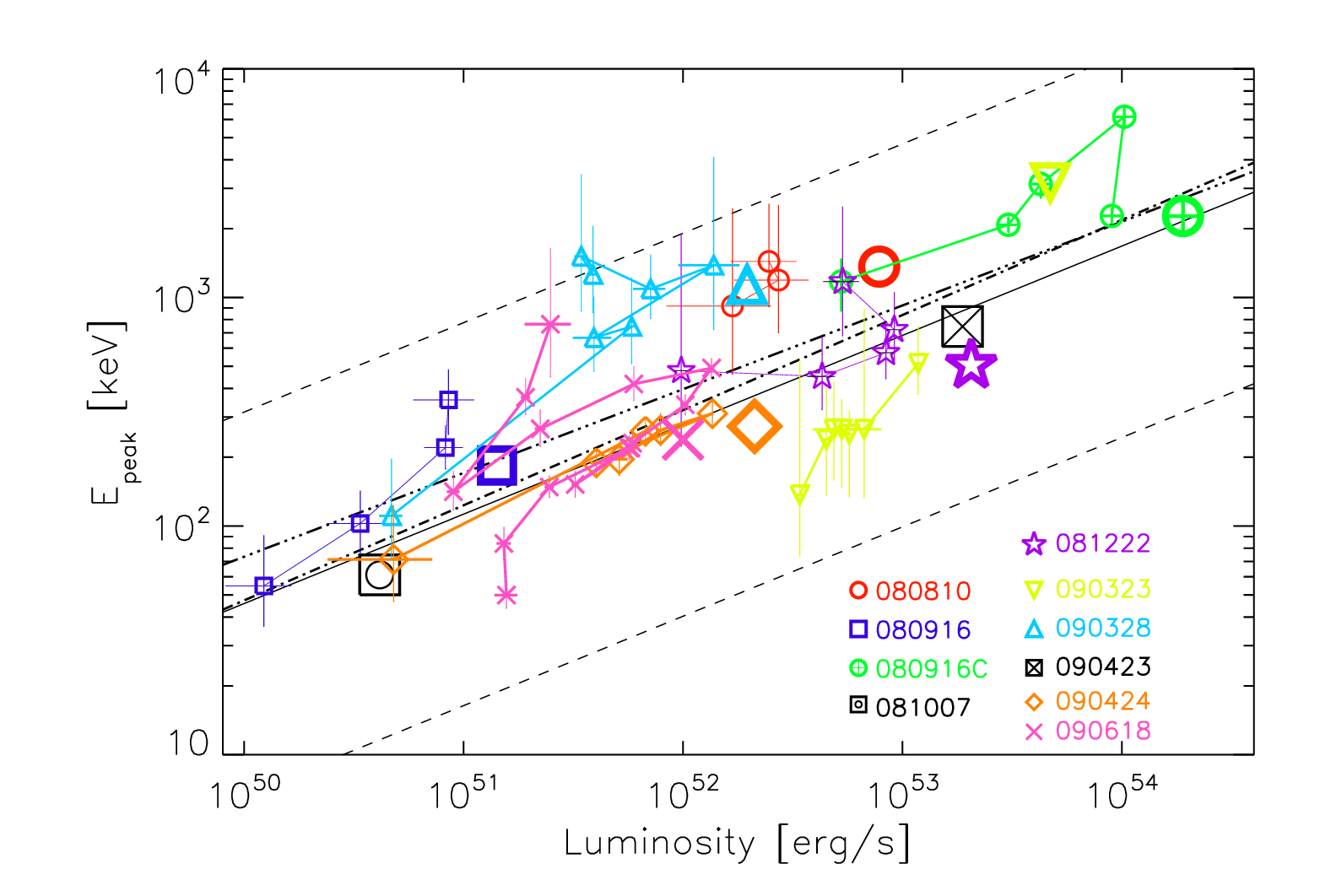}
\includegraphics[width=2.8in, height=2.7in]{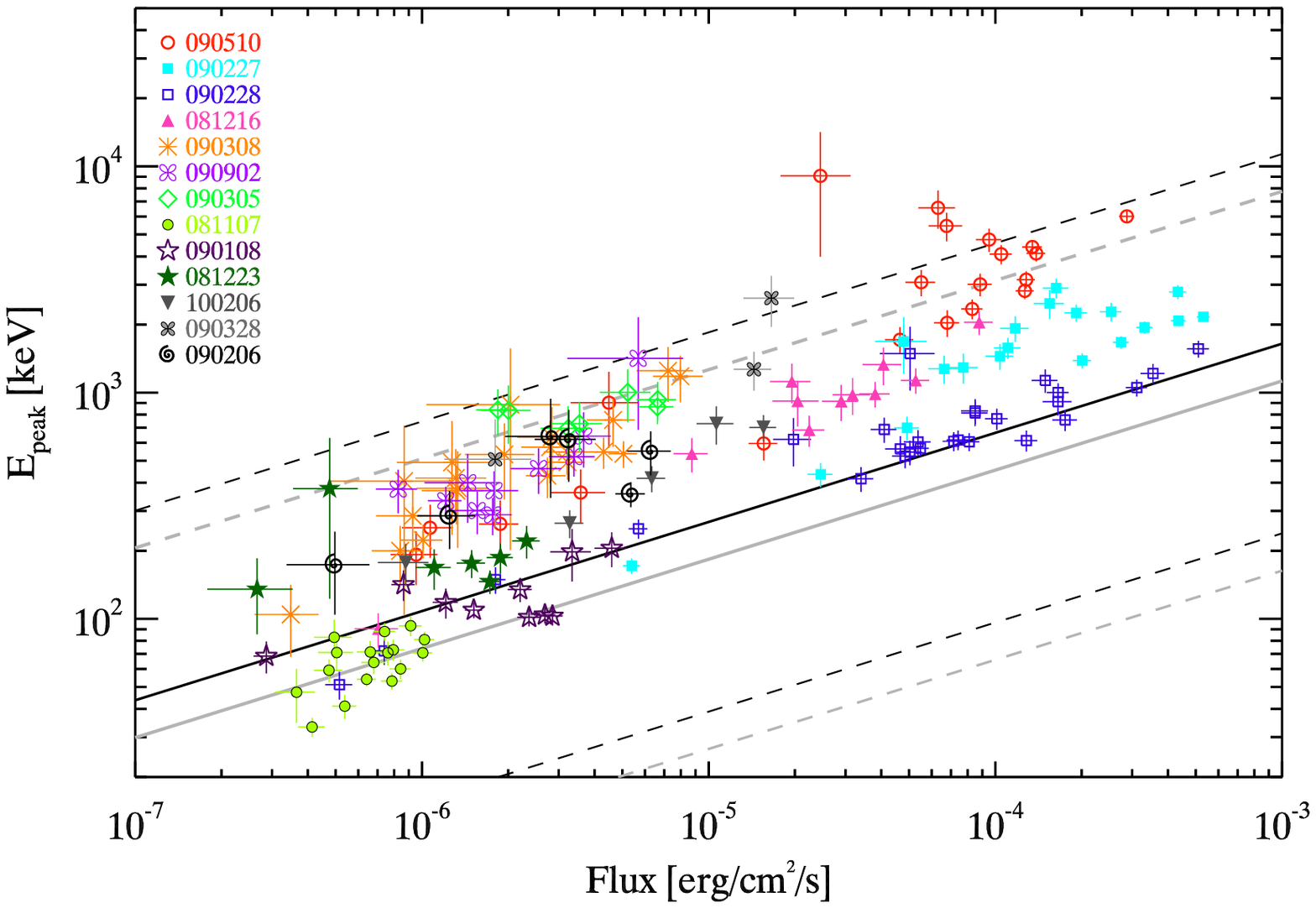} 
\vspace*{-0.6 cm}
\caption{Left: time--resolved \yone\ correlation of 10 \fe\ {\bf long} GRBs with known redshift (shown with open red circles in Fig.~\ref{fig1}). 
Small symbols show the evolution of \ep\ vs \liso. Large symbols are the location in the \yone\ plane of the corresponding bursts when the 
time--integrated spectra are considered (for GRB 081007 and GRB 090423, only the time--integrated spectrum is available). 
The solid and dotted lines represent the \yone\ correlation and its 3$\sigma$ scatter, respectively, as obtained with the pre--Fermi GRBs (filled grey circles 
in Fig.~\ref{fig1}. The dot--dashed line represents the Þt to the 10 Fermi GRBs (time--integrated) and the triple--dot--dashed line is the Þt to the 
51 time--resolved spectra (Ghirlanda et al. 2010). Right: time--resolved correlation between the peak energy \ep\ and the flux of the 163 time resolved spectra of the 13 {\bf short} \fe\ GRBs analyzed in Ghirlanda et al. 2010a. Different symbols/colors correspond to different bursts (as shown in the legend). For all but one short GRBs (090510 - open blue square in Fig.~\ref{fig1}) the redshift is not known, this is why their spectral evolution is represented in the observer frame.  The solid line (dashed lines) is the \yone\ correlation (and its 3$\sigma$ scatter) of long GRBs transformed in the observer frame assuming $z=1$ ($z=0.5$ for the grey solid and dotted lines).
 }
\label{fig2}
\end{figure}

We have analyzed the time resolved spectra of 10 long GRBs with measured redshifts detected by \fe\ up to July 2009. Their spectral evolution shows that the
peak energy tracks the flux and a strong time--resolved \yonet\ correlation is present within individual GRBs (Fig.~\ref{fig2} left panel). The time resolved \yonet\ correlation is similar to the time--integrated one (solid line in Fig.~\ref{fig2} left panel). The same analysis was applied to short GRBs detected by \fe. Except for GRB 090510 at $z=0.9$ (open blue square in Fig.~\ref{fig1}), the redshifts of the other \fe\ short bursts is not measured. However, we can still study the correlation between the observer frame peak energy and the bolometric flux (Fig.~\ref{fig2} right panel). Also in short bursts there is strong hardness--intensity correlation. It is consistent with the \yone\  transformed in the observer frame by assuming a typical redshift $z=0.5$ (grey solid line in Fig.~\ref{fig2}) or $z=1$ (black solid line in Fig.\ref{fig2}). 

\begin{figure}
\vspace*{-0.5 cm}
\hspace*{-0.8 cm}
\includegraphics[width=5.5in, height=2.8in]{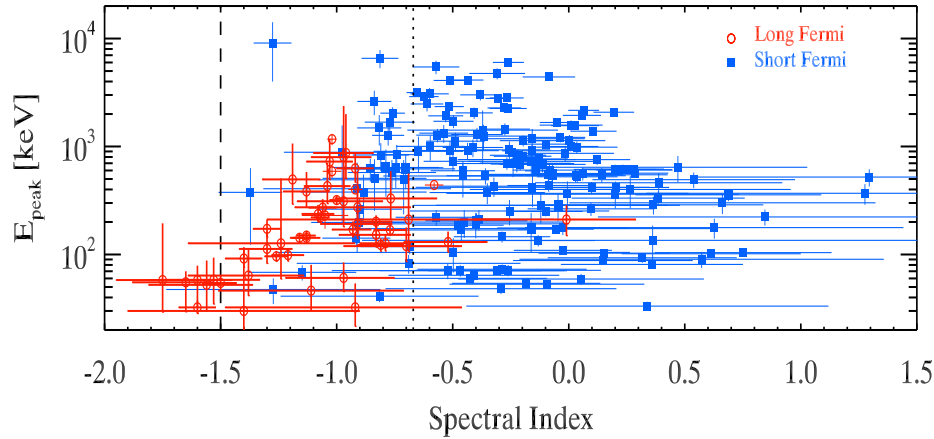}
\vspace*{-0.6 cm}
\caption{Observer frame peak energy versus low energy spectral index of the time resolved spectra of the 10 long GRBs (open red circles) 
and 14 short GRBs detected by \fe\ and analyzed in Ghirlanda et al. 2010 and Ghirlanda et al. 2010a, respectively. The dotted (dashed) 
vertical line is the synchrotron limit (with cooling).}
\label{fig3}
\end{figure}

In order to verify that the time resolved hardness--intensity correlation found is not induced by the spectral model used to fit the time resolved spectra, we show in Fig.~\ref{fig3} that there is no correlation between the low energy spectral index $\alpha$ and the peak energy \ep\ in both short (filled blue squares) and long GRBs (open red circles). Fig.~\ref{fig3} also shows that short GRBs have harder low energy spectral index than long events as a confirmation of what already found from the comparison of their time integrated spectra (Ghirlanda et al. 2004, 2009 - where it was instead shown that the peak energies are similar in short and long events). 

Recently, Guirec et al. (2010), by studying the spectral evolution of the three short brightest \fe\ GRBs, find that they have larger \ep\ than long ones, claiming that this (apparently) contradicts what found in Ghirlanda et al. (2004, 2009). Not surprisingly, this is simply due to the existence of a correlation between \ep\ and the flux in both short and long GRBs. Indeed, the three brightest short GRBs analyzed by Giuriec et al. 2010, which are also present in our sample of 14 events, reach values of \ep\ significantly harder than long events but they are also significantly brighter. For the same fluxes short and long GRBs have similar \ep.

\section{Conclusions}

All  the spectral--energy correlations in GRBs have been derived considering the time--integrated GRB spectral properties. The spectral analysis of 10 \fe\ GRBs with measured $z$ and 14 short \fe\ GRBs reveals the existence of a time--resolved  \yonet\ within individual bursts. The evolutionary tracks lie in the upper part of the \yone\ correlation (Fig.~\ref{fig2}). This could be caused by a systematic underestimate (overestimate) of the luminosity (peak energy) in time--resolved spectra with respect to time--integrated spectra because the former are more frequently fitted with a CPL model that lacks the high--energy power law component (Band function) typically fitted to time--integrated spectra.  The existence of a \yonet\ correlation within individual GRBs, consistent with the \yone\ correlation defined by time--integrated spectra, is the strongest argument in favour of a physical origin of this correlation (or the other way round, the strongest argument against selection effects). If the origin of the spectral energy correlations is to be found in the emission mechanism, then it should be the same for short and long GRBs. 

\begin{acknowledgments}
I am grateful to G.Ghisellini, L. Nava, D. Burlon, M. Nardini, A. Celotti for useful discussions. 
ASI is thanked for grant I/088/06/0.  A PRIN-INAF grant is acknowledged for 
funding. This research made use of the \fermi--Gamma Burst Monitor data 
publicly available via the NASA-HEASARC data center.  
\end{acknowledgments}

\end{document}